%% file: main.tex
\documentclass[conference,compsoc]{IEEEtran}

\usepackage{xspace}
\usepackage{graphicx}
\usepackage{enumitem}
\usepackage{booktabs}
\usepackage{multirow}
\usepackage[binary-units=true]{siunitx}
\usepackage{amsfonts,amsthm,amsmath}
\usepackage{subcaption}
\usepackage{latexsym}       
\usepackage{wasysym}        
\usepackage{tcolorbox}
\usepackage{tikz}
\usepackage{makecell}

\usepackage{xcolor}
\usepackage{siunitx}
\usepackage[hyphens]{url}
\usepackage{framed}
\usepackage{bm}
\usepackage[ruled,noend]{algorithm2e}
\usepackage{algpseudocode}
\usepackage{tabularx}
\usepackage{pifont}
\usepackage{textcomp}

\renewcommand{\paragraph}[1]{\vspace{5pt}\noindent\textbf{#1.}\xspace}

\input{macros}

\begin{document}

\title{Supporting Human Raters with the Detection\\of Harmful Content using Large Language Models}

\author{Kurt Thomas$^1$ \quad 
Patrick Gage Kelley$^1$ \quad 
David Tao$^1$ \quad 
Sarah Meiklejohn$^1$ \quad 
Owen Vallis$^1$\\
Shunwen Tan$^1$ \quad
Blaž Bratanič$^2$ \quad
Felipe Tiengo Ferreira$^2$ \quad
Vijay Kumar Eranti$^1$ \quad
Elie Bursztein$^1$ \vspace{5pt}\\
\emph{$^1$Google \qquad $^2$Google DeepMind}
}

\maketitle

\input{00_abstract}
\input{00a_intro}

\input{01_related}

\input{02_design}

\input{03_methodology}

\input{04_experiments}

\input{04a_patterns}
\input{05_deployment}

\input{06_discussion}

\input{07_conclusions}

{\footnotesize
\bibliographystyle{abbrv}
\interlinepenalty=10000
\bibliography{references}
}
\input{08_appendix}

\end{document}

%% file: macros.tex
\newcommand{\eg}{e.g., }

\newcommand{\ie}{i.e., }
\newcommand{\etal}{et al.\@\xspace}

\usepackage[utf8]{inputenc}

\newcommand{\fullcirc}{$\newmoon$}
\newcommand{\emptycirc}{$\times$}

\usepackage{array}
\newcolumntype{L}[1]{>{\raggedright\let\newline\\\arraybackslash\hspace{0pt}}m{#1}}
\newcolumntype{C}[1]{>{\centering\let\newline\\\arraybackslash\hspace{0pt}}m{#1}}
\newcolumntype{R}[1]{>{\raggedleft\let\newline\\\arraybackslash\hspace{0pt}}m{#1}}

\usepackage{array}
\newcolumntype{P}[1]{>{\raggedleft\arraybackslash}p{#1}}

\newcommand{\vertexsmall}{{\small\sffamily text-bison}\xspace}
\newcommand{\vertexlarge}{{\small\sffamily text-unicorn}\xspace}
 
\newcommand{\vertexsmallshort}{{\small\sffamily bison}\xspace}
\newcommand{\vertexlargeshort}{{\small\sffamily unicorn}\xspace}

\newcommand{\Google}{Google\xspace}

\newcommand{\boldlab}{\bm{$\langle$}}
\newcommand{\boldrab}{\bm{$\rangle$}}

%% file: 00_abstract.tex
\begin{abstract}
In this paper, we explore the feasibility of leveraging large language models (LLMs) to automate or otherwise assist human raters with identifying harmful content including hate speech, harassment, violent extremism, and election misinformation. Using a dataset of 50,000 comments, we demonstrate that LLMs can achieve 90\% accuracy when compared to human verdicts. We explore how to best leverage these capabilities, proposing five design patterns that integrate LLMs with human rating, such as pre-filtering non-violative content, detecting potential errors in human rating, or surfacing critical context to support human rating. We outline how to support all of these design patterns using a single, optimized prompt. Beyond these synthetic experiments, we share how piloting our proposed techniques in a real-world review queue yielded a 41.5\% improvement in optimizing available human rater capacity, and a 9--11\% increase (absolute) in precision and recall for detecting violative content.
\end{abstract}

%% file: 00a_intro.tex
\vspace{12pt}
\noindent\fbox{%
    \parbox{0.98\columnwidth}{%
        This paper includes descriptions and quotes of harmful content. We have redacted this content to minimize harm; take care of yourself when engaging with this material.
    }%
}

\section{Introduction}
Protecting users from abuse has expanded in its purpose---from an original focus on scams, phishing, and malware---to now encompass a spectrum of \emph{harmful content}. Threats include hate and harassment that attempt to silence online voices~\cite{thomas2021sok}; violent extremism that glorifies terrorist attacks~\cite{edwards2013pathways}; and misinformation and disinformation that diminishes trust in institutions (\eg vaccines approved by the medical establishment or the integrity of fair elections)~\cite{kennedy2022repeat, loomba2021measuring}. In practice, robustly defining harmful content requires a multitude of specialized policies, the nuances and enforcement of which differ from platform to platform and by country or locality~\cite{singhal2023sok, pater2016characterizations, gorwa2020algorithmic}.

Platforms have responded to harmful content with a defense-in-depth approach. Protections include content classifiers and hash databases that detect \emph{violative content}, such as Jigsaw's Perspective API~\cite{lees2022new}, the GIFCT terrorist image database~\cite{gifct}, and recent APIs from Google and Microsoft~\cite{google-content-safety, microsoft-content-safety}. User reporting serves to both surface emerging threats as well as to call attention to content missed by detection systems~\cite{crawford2016flag}. And appeals serve to safeguard against \emph{non-violative content} being inaccurately flagged and removed~\cite{myers2018censored, vaccaro2020end}.

Ultimately, training data and decisions for all of these systems and policies hinges on \emph{human raters} who bring in-context, human expertise. This poses an ongoing challenge for a variety of reasons. Expertise is a scarce resource compared to the scale of user content generated daily. Policy clauses (and interpretations) can rapidly change in response to new product features, regulations, and emerging threats (\eg evolution in hate speech~\cite{zannettou2018origins}). Decisions can require regional knowledge pertaining to languages, cultures, and memes~\cite{zhu2021self, zannettou2018origins}. Finally, exposure to harmful content incurs a burden on those involved~\cite{steiger2021psychological}. Streamlining or improving human decision making can thus help with the scale and efficacy of protecting users from harmful content, while also ensuring more consistent and fair enforcement~\cite{ma2022m}.

In this paper, we explore how to harness a sea-change moment with the emergence of large language models (LLMs) to optimize what content receives a human rater's consideration and how human raters arrive at a policy decision. Specifically, we examine:

\begin{enumerate}[label={\bfseries RQ\arabic*:},leftmargin=1cm, topsep=0pt,itemsep=0ex,partopsep=1ex,parsep=1ex]

    \item What prompt engineering practices, without fine-tuning, maximize an LLM's ability to interpret real-world policies for multiple types of harmful content?
    
    \item What design patterns---such as filtering non-violative content, rapidly escalating violative content, or surfacing critical context to human raters---best utilize the current capabilities of LLMs? 
    
    \item What is the real-world impact of deploying an LLM to assist human raters with detecting harmful content?
    
\end{enumerate}

To answer these questions, we leverage a dataset of 40,000 text comments written in English that human raters at \Google identified as violating its hate speech, harassment, violent extremism, or election misinformation policies~\cite{google-policies}; as well as 10,000 comments that human raters determined did not violate any of Google's policies. We experiment with a variety of prompting strategies to classify each comment (\eg retrieval augmented generation~\cite{lewis2020retrieval} and chain of thought~\cite{wei2022chain}) and evaluate each prompt variant on Google's PaLM 2 \vertexsmall and \vertexlarge models~\cite{google-vertex-ai}.

We find that LLMs can achieve accurate policy decisions
compared to human raters, with our best prompt on \vertexlarge achieving 98.7\% accuracy for election misinformation, 91.1\% accuracy for hate speech, 91.1\% accuracy for violent extremism, and 90.1\% accuracy for harassment. At the same time, we show that a single prompting strategy is flexible enough to support a variety of design patterns. These include pre-filtering likely non-violative content, rapidly escalating violative content, detecting potential errors in human ratings, and surfacing critical context to human raters.

We demonstrate the real-world feasibility of our proposed approach on a review queue at \Google that covers 20+ harmful content policies. We show that just a single prompt, encompassing every policy, achieves 95.1\% recall of violative content, thus allowing automated decisions for 41.5\% of the queue so human raters can focus their time on borderline content---rather than clearly benign content. Furthermore, in an experiment where LLMs assist human raters with detection by highlighting relevant text passages, human rating precision and recall both increase by 9--11\% (absolute). Taken as a whole, our results highlight how LLMs represent a promising direction for scaling human expertise for trust and safety, which can reduce the time to decision making, reduce burden on human raters, ensure consistency, and in some cases improve overall accuracy.

%% file: 01_related.tex
\section{Related work}\label{sec:related}

\paragraph{Defining and understanding harmful content} The scale and scope of policies that govern harmful content have evolved along with the global user bases of online platforms~\cite{Gorwa2020-nx}. Researchers have explored some of the most pressing categories of harmful content, including hate speech and harassment~\cite{mondal2017measurement, silva2016analyzing, thomas2021sok}, violent extremism~\cite{edwards2013pathways}, misinformation~\cite{kennedy2022repeat, loomba2021measuring}, and child safety~\cite{bursztein2019rethinking,quayle2020prevention}; as well as the idiosyncrasies of policies across platforms~\cite{pater2016characterizations, singhal2023sok}. Other works have looked to systematize the space according to the harms different content might incur~\cite{scheuerman2021framework} or the viability of interventions~\cite{singhal2023sok}. 

Additionally, major platforms are shifting towards greater transparency, releasing more detailed information on the specific policies they use to moderate content, partially due to requirements in the Digital Services Act (DSA)~\cite{appelman2021using}; examples include the Meta Transparency Center~\cite{fb-transparency} and Google Transparency Center~\cite{google-policies}. This offers researchers better visibility into how platforms make moderation decisions, and what content is considered violative.

\paragraph{Detecting harmful content} Platforms and researchers have developed a plethora of detection strategies for identifying harmful content, most often in the form of content classifiers for text~\cite{Davidson2017-cx,Zampieri2019-gz,fortuna2018survey,lees2022new,google-content-safety,microsoft-content-safety,Salminen2018-hm} or hash matching solutions for imagery~\cite{gifct, bursztein2019rethinking}. Beyond algorithmic solutions, platforms also rely on user reporting \cite{crawford2016flag}, and verdict appeals \cite{myers2018censored,vaccaro2020end}. For a more comprehensive overview, see Singhal \etal~\cite{singhal2023sok}.

\paragraph{Datasets of harmful content} Public datasets of harmful content are currently scarce and often rely on crowd-sourced ratings, rather than ratings from trained experts. The datasets that exist primarily focus on a spectrum of hate speech and harassment~\cite{jigsaw-kaggle, Mollas2020-eu, Zampieri2019-gz, kumar2021designing}, and to a lesser extent, misinformation~\cite{nielsen2022mumin}. Previous studies have shown these datasets contain biases~\cite{Davidson2019-cr} and that the labels produced by crowd-sourced workers contain significant disagreement~\cite{gordon2022jury, kumar2021designing}. As such, we opted to use a non-public corpus of harmful content that reflects the reality of content moderation on a major platform backed by raters trained on real policies.

\paragraph{LLMs as classifiers} Since the emergence of powerful LLMs, a variety of researchers have investigated the zero-shot and k-shot capabilities of available models. For example, Chiu \etal used GPT-3 to detect racist and sexist language using k-shot prompts \cite{Chiu2021-tv}. However, they cautioned that smaller LLMs might lack the sensitivity to differentiate between terms associated with targeted or protected groups and their context. Huang \etal conducted a similar experiment, targeting implicit hate speech with GPT-3~\cite{Huang2023-rl}. Rather than prompting, He \etal explored fine-tuning models for toxicity detection, achieving a 10\% performance gain over existing baselines~\cite{he2023you}. Bai \etal proposed a model alignment approach to enhance LLM sensitivity to harmful content by providing specifications and principles \cite{Bai2022-ho}. This enabled the model to critique its output and improve its awareness of offensive content. Similarly, Vishwamitra \etal explored identifying derogatory terms and generalizing detection via chain of thought using a BERT model~\cite{vishwamitra2023moderating}. Closest to our work, Weng \etal recently explored whether GPT-4 could achieve similar accuracy to human raters for a variety of moderation tasks~\cite{Weng2023-mz}. Via a blog post, they reported that LLM performance approached that of reviewers with limited training, but fell short of expert reviewers. However, their post shared no experimental details (\eg setup, prompt strategies, or design strategies). In short, while others have considered the use of LLMs for content moderation, we depart from the conventional design choice of LLMs as automatic content classifiers and focus instead on enhancing and maximizing the expertise of human raters.

%% file: 02_design.tex
\section{Collaborative design patterns}\label{sec:design}
\begin{figure}
    \centering
    \includegraphics{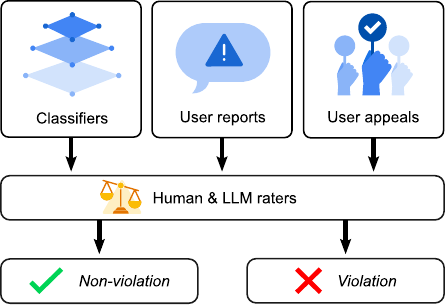}
    \caption{Platforms rely on human raters to label data for supervised abuse classifiers, to triage user reports of suspected policy violations, and to determine the validity of user appeals that claim content is non-violative. We envision an LLM agent that optimizes which content gets sent to human raters and assists human raters in arriving at decisions.}
    \label{fig:review_flow}
\end{figure}

We begin by outlining how LLMs can assist human experts with detecting harmful content in a variety of policy contexts (\eg hate speech, misinformation). In practice, human ratings are critical to generating training and testing data for supervised abuse classifiers, triaging user reports of policy violations, and handling appeals of suspected non-violations as shown in Figure~\ref{fig:review_flow}. Due to the asymmetry in the scale between the volume of user-generated content requiring rating decisions and available human expertise for billion-user platforms, human raters are a scarce resource. Our vision is that LLMs can automate clear decisions around harmful content that do not require subtle human expertise and assist human raters in their deliberation process on the remaining content. Here, we enumerate five such collaborative design patterns along with their terminology and associated performance metrics.

\subsection{Terminology}

\paragraph{Content} \emph{Content} or \emph{object} refers to any user-generated data, such as comments, reviews, posts, and more. For the purposes of this work, we focus on text-based content, though emergent multi-modal LLMs would also allow for the classification of images, videos, audio, and other content.

\paragraph{Policy} A \emph{policy} is any set of natural language rules that a platform uses to determine whether or not content should be prohibited for safety reasons. Examples include policies against scams, hate speech, and misinformation.

\paragraph{Verdict} We refer to the outcome of a policy determination as a \emph{verdict}. A verdict may deem the content \emph{violative}, thus triggering an escalation such as the removal of the content from the platform, or \emph{non-violative} resulting in no further action. The choice of escalation is context dependent and outside the scope of this work.

\paragraph{Human rater} We assume the existence of a \emph{human rater}: a human expert responsible for making policy decisions. In practice, their expertise---and thus accuracy---may vary due to ambiguity, necessary local context, a spectrum of familiarity with the policy, rapidly changing policies, or complex cases with competing policy interest (\eg free speech concerns vs.\ potential for harm).

\paragraph{LLM rater} We refer to any LLM agent that assists with policy determinations as an \emph{LLM rater}. As we discuss shortly, the assistance this LLM rater provides can vary in purpose: it can automate decision making, identify errors produced by human raters, or augment the decision making process of human raters.

\paragraph{Rater queue} We assume the existence of a \emph{rater queue} that includes content---or objects---that should be evaluated against one or more policies.

\begin{figure*}[t]
    \centering
    \includegraphics[width=0.95\textwidth]{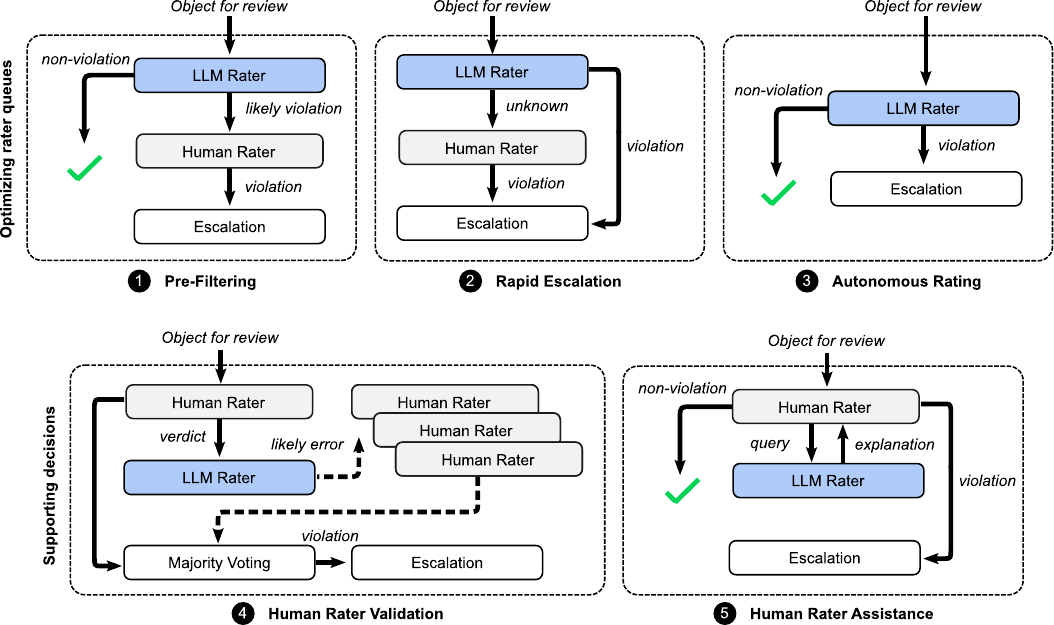}
    \caption{Design patterns for using an LLM to assist human raters. We separate these into designs that attempt to optimize which content is sent to human raters (\ding{182}, \ding{183}, \ding{184}) and designs that attempt to improve the accuracy of human raters (\ding{185}, \ding{186}).}
    \label{fig:llm_modes}
\end{figure*}

\subsection{Performance metrics}
\label{sec:performance}
We assess the performance of LLM raters according to improvements in scaling, latency, and accuracy. While there are other potential benefits of LLM raters---such as improving the well-being of human raters analyzing harmful content~\cite{steiger2021psychological}---these are highly context dependent and thus outside the scope of our study (beyond simply reducing the overall number of reviews performed, as discussed below).

\newcommand{\Mcost}{\textbf{M1}}
\newcommand{\Mlatency}{\textbf{M2}}
\newcommand{\Mfn}{\textbf{M3}}
\newcommand{\Mfp}{\textbf{M4}}

\paragraph{M1 -- Scaling} Human rating is a finite and expensive resource. Assuming that the cost of querying an LLM is less than that of querying a human rater, one can scale the volume of reviews beyond what was previously feasible by using LLMs to prioritize what content requires the expertise of human raters, meanwhile automating the bulk of rote decisions. Any scaling in practice is context dependent. As a rough feasibility estimate, Google Cloud charges \$0.0005 per thousand characters of text input and output to PaLM 2;\footnote{{\scriptsize\url{https://cloud.google.com/vertex-ai/pricing\#generative\_ai\_models}}}  Microsoft Azure charges \$0.001 per thousand input tokens and \$0.002 per thousand output tokens for GPT-3.5.\footnote{\scriptsize\url{https://azure.microsoft.com/en-us/pricing/details/cognitive-services/openai-service/}} While there is limited public information on human rater salaries, data annotation services such as SageMaker charge \$0.08 per object\footnote{\scriptsize\url{https://aws.amazon.com/sagemaker/data-labeling/pricing/}}---on the order of 100x compared to LLMs. 

\paragraph{M2 -- Reducing decision latency} Human rating incurs a decision delay $t_m$ on the order of seconds to minutes (not including time spent in a queue). Given the latency of querying an LLM as $t_{m'}$ for the same object, we can estimate the reduction in decision latency as $t_m - t_{m'}$.

\paragraph{M3 -- Reducing false negatives} Human rating, despite relying on experts, is still subject to errors. If an LLM rater can outperform a human rater at identifying violations, this can lead to an overall reduction in false negatives (\eg incorrectly missed violations) and potentially more consistent enforcement.

\paragraph{M4 -- Reducing false positives} As above, if an LLM rater can outperform a human rater at identifying non-violations, this can lead to an overall reduction in false positives (\eg incorrectly escalated violations) and potentially more consistent enforcement.

\subsection{Modes of operation}\label{sec:modes}
We envision multiple modes of operating an LLM rater to improve content moderation as shown in Figure~\ref{fig:llm_modes}. Real world deployments will likely rely on a combination of some or all of these versions of LLM raters. We discuss the relative merits of each mode below.

\paragraph{\ding{182} Pre-filtering [\Mcost, \Mlatency]} For rater queues that are heavily skewed towards non-violative content (\eg noisy user reports or rare incidents), an LLM rater can operate as a pre-filter before sending content to a human rater. Here, the LLM rater needs to achieve a minimum threshold $T_r$ of \emph{recall}\footnote{We use the standard machine learning definition of recall---or sensitivity---which is correctly identified violative samples divided by the total number of violative samples: $TP/P$.} to ensure that a majority of violative content is forwarded to a human rater, while also achieving a high degree of \emph{specificity}\footnote{We use the standard machine learning definition of specificity, which is the total number of correctly identified non-violative samples divided by the total number of non-violative samples: $TN/N$.} to ultimately dequeue non-violative content. Pre-filtering serves to free up raters to review more likely violative content while increasing the total volume of content reviewed (\Mcost). It may also potentially expedite decision making if the rater queue is non-saturated after filtering (\Mlatency).

\paragraph{\ding{183} Rapid escalation [\Mcost, \Mlatency, \Mfn]} For rater queues that may contain high-severity or emerging threats that requires immediate escalation, an LLM rater can operate as a fast-path escalation, entirely bypassing any human rater (or putting objects at the head of a queue). Doing so requires that the LLM rater achieves a minimum threshold $T_p$ of \emph{precision}\footnote{We use the standard machine learning definition of precision, which is correctly identified violative content divided by all content flagged as violative: $TP / (TP + FP)$.} and a high degree of recall to capture as much high-severity content as possible. Such a system might be deployed temporarily in response to an incident despite low precision (with a higher risk of appeals), or long-term if the precision is sufficiently high. This escalation architecture serves to expedite decision making (\Mlatency) and potentially avoid false negatives that a human rater may make (\Mfn) if the LLM rater can outperform a human rater in terms of precision and sensitivity. For sufficiently large volumes of violations, it may also allow additional scaling (\Mcost).

\begin{table*}[t]
    \input{tables/datasets}
    \caption{Stratified sample of English comments identified by human raters at \Google for violating one of four distinct policies, as well as a sample of non-violative comments. Examples have been paraphrased for privacy and redacted to minimize repeating harmful language.}
    \label{table:dataset}
\end{table*}

\paragraph{\ding{184} Autonomous rating [\Mcost, \Mlatency, \Mfn, \Mfp]} For some policies, it may be possible for an LLM rater to achieve equivalent---or better---performance to human raters, thus enabling the full automation of a rater queue. This represents a more difficult challenge as the LLM must achieve both a minimum threshold for precision $T_p$ and recall $T_r$, with both thresholds ideally exceeding the performance of human raters. As a safeguard, any appeals of LLM rater verdicts by users should not be sent to the same LLM rater. This operation mode has the potential to improve all performance dimensions of the rater queue.

\paragraph{\ding{185} Human rater validation [\Mfn, \Mfp]} Rather than automating (some) decisions in a rater queue, an LLM rater can serve as a quality control system. Mechanistically, an LLM rater and human rater both rate the same content. If they disagree on the verdict---and the LLM rater has sufficient confidence---the same content can be sent to one or more additional human raters. The final verdict then consists of a majority vote across human raters. This quality check can bias towards identifying errant non-violative verdicts (\Mfn) or alternatively errant violative verdicts (\Mfp). Configuration of the quality check requires balancing the total number of additional human ratings an LLM rater can trigger versus the lift in accuracy.

\paragraph{\ding{186} Human rater assistance [\Mlatency, \Mfn, \Mfp]} An LLM rater can also help human raters understand the context of potential violative content. During rating, a human rater might engage with an LLM rater to receive an explanation of why content is violative or not---such as highlighting specific passages of content that may violate a policy. This augmentation can help expedite rating (\Mlatency) while also drawing attention to critical context that might otherwise be missed, improving overall accuracy (\Mfn, \Mfp). However, there is also the risk of biasing human raters to agree with the LLM rater.

%% file: tables/datasets.tex
\centering
\small
\begin{tabularx}{\textwidth}{l|X|r}
\toprule
\bf Policy type & \bf Example comments & \bf Sample size \\
\midrule
    \multirow{2}{*}{Hate Speech}& \it ``Because every single [redacted] steals and commits crimes. It’s their culture.'' & \multirow{2}{*}{10,000} \\
    & \it ``Sadly due to mass immigration of low IQ savages [redacted] is a cesspit.'' & \\
    \midrule
        \multirow{2}{*}{Violent Extremism} & \it ``Lethal attachments are a dime a dozen.  Ill be a mass shooter in no time.'' & \multirow{2}{*}{10,000} \\
        & \it ``I hope more people like [Christchurch shooter] emerge and prove them right.'' & \\
    \midrule
    \multirow{2}{*}{Harassment} &\it ``Go find the tallest building then jump bro its literally not that serious.'' & \multirow{2}{*}{10,000} \\
    & \it ``Somebody SHUT that nasal, whiny little [redacted] up!'' & \\
    \midrule
    \multirow{2}{*}{Election Misinformation} & \it``Covid was a cover to remove [redacted] through fraudulent mail in voting.'' &  \multirow{2}{*}{10,000} \\
    & \it ``YOU ARE AWARE ARN,T YOU THE VOTING MACHINES WERE RIGGED.'' & \\
    \midrule
    \multirow{2}{*}{Non-violative} &\it ``Bro you better apologize, you messed with wrong people.'' & \multirow{2}{*}{10,000} \\
    & \it ``Average IQ in this comment section is 90.'' & \\
    \bottomrule
\end{tabularx}

%% file: 03_methodology.tex
\section{Dataset}\label{sec:methodology}
In order to explore the feasibility of LLM raters and our collaborative design patterns, we obtain a real-world dataset of 50,000 human-rated violative and non-violative comments for a variety of types of harmful content. We discuss this dataset, our approach to validation, and limitations that stem from our approach.

\subsection{Policy violations and non-violations} 
Our dataset consists of a stratified sample of 40,000 unique, user-generated comments that human policy raters at \Google identified as violative for one of four distinct policies: hate speech, violent extremism, harassment, and election misinformation. This dataset is ecologically valid, representing the reality of human ratings at-scale for diverse user bases. We focus on these policies as it allows us to assess the generalizability of an LLM rater to multiple safety concepts, while balancing the overhead of querying LLMs with tens of thousands of comments per experimental design. We provide a sample of violative comments for each policy in Table~\ref{table:dataset}. Additionally, we have a sample of 10,000 unique, user-generated comments that human raters determined were non-violative of all \Google's policies.\footnote{In practice, \Google's policies are more expansive than the four we evaluate. As such, non-violative content excludes scams and many other forms of harmful content. As we show, many of these non-violative comments border policies in some way, leading to the potential for false positives.} All comments are in English, which is one of the most popular languages on \Google. For each sample, we have the text of the comment and a binary label (\eg 1 if violative, 0 if non-violative). Each policy strata reflects a randomized sample drawn over a three month period from August 2023--October 2023, with the exception of election misinformation which was drawn from October 2022--October 2023. For all our experiments, we use a \emph{balanced} dataset of violative samples and non-violative samples per policy to evaluate the capabilities and tradeoffs of LLM rating. Later in Section~\ref{sec:deployment}, we discuss the real-world pilot of our proposed techniques on live rating queues on the natural distribution of violations and non-violations.

\subsection{Validation}\label{sec:methods_validation} For the purposes of our study, we treat the labels from human policy raters as \emph{ground truth}. In practice, these labels may include some errors. As an assessment of data quality, we randomly sampled 500 comments, stratifying this selection to include 100 comments per each policy type in our study---including non-violations. Two researchers independently labeled each comment as violative or non-violative according to the public policy text, with a third researcher helping to resolve all disagreement. We also allowed for a third label of ``missing context'', in the event the three researchers could not render a judgement without the comment's surrounding context (\eg ``They look like aliens'' is ambiguous for who ``they'' refers to).\footnote{Unlike our research team, policy raters had access to the surrounding context (\eg the page the comment appeared on, other comments in a conversation) when making their decision.}

We present the rate of agreement and disagreement between the original expert labels and validation labels in Table~\ref{table:dataset_validation}. For policy violations, we find just 0--2\% of comments may be non-violations. Another 0--8\% of comments lacked sufficient context for us to make an absolute determination. For example, ``this mental illness stuff is getting out of hand fr'' was flagged by an expert as hate speech (likely due to being transphobic and claiming transgender individuals have a mental illness), but our validation lacks the surrounding context, so we cautiously assigned a label of missing context. The remaining 90--98\% of comments were clearly violative. For non-violative content, we find 8\% of comments may violate one of the four policies in our study. Another 15\% might be construed as violative depending on the missing context. For example, ``Shoot that dirtbag horse!'' might be threatening violence, or be about a video game, so we assigned a label of missing context. The remaining 77\% of comments were clearly non-violative.

We caution that the researchers involved in validation are not trained policy raters. It is clear from our validation that many non-violative comments are offensive or incendiary, but not necessarily to the level of a policy violation. Through this lens, we argue it is methodologically sound to treat the original expert labels as ground truth. The consequence on our measurements is that an LLM may have an upper bound of 98--100\% for recall, and an upper bound of 92\% for precision, if validation labels (\eg ``Disagree'' in Table~\ref{table:dataset_validation}) are in practice more accurate than the original expert labels. 

\begin{table}[t]
    \centering
    \input{tables/dataset_validation}
    \caption{Results of our independent validation of 500 ground truth labels. We find 0--2\% of comments flagged by human policy raters as violations may be non-violations. Another 8\% of comments deemed non-violations may be violations.}
    \label{table:dataset_validation}
\end{table}

\subsection{Ethics and Limitations}
Our methodology incurs a number of limitations. Our dataset is limited to four types of harmful content. While these types cover some of the most pressing emerging threats, enforcement may differ from platform to platform, even if policy language may be similar. Likewise, the distribution of comments from \Google may not generalize to other platforms. These subtleties may result in variable performance when evaluated on other datasets. Finally, in the event of invalid ground truth labels, we may inaccurately estimate the performance that LLMs can achieve. We note the researchers in this study never had access to any identifying information associated with comments (\eg usernames or account activities), and that all comments were posted publicly.

%% file: tables/dataset_validation.tex
\small
\begin{tabularx}{\columnwidth}{X|P{1cm}|P{1cm}|P{1cm}}
\toprule
\bf Policy & \rotatebox{0}{\bf Agree} & \rotatebox{0}{\makecell{\bf  Missing \\ \bf context}} & \rotatebox{0}{\bf Disagree} \\
\midrule
Hate Speech &	90\% & 8\% & 2\%\\
Violent Extremism &	96\% & 4\% & 0\%\\
Harassment &	91\% & 7\% & 2\%\\
Election Misinformation &	98\% & 0\% & 2\%\\
Non-violative &	77\% & 15\% & 8\%\\
\bottomrule
\end{tabularx}

%% file: 04_experiments.tex
\section{Building LLM raters}\label{sec:experiments}

We conduct an ablation study to determine how different choices of prompts and models affect the performance of an LLM rater in terms of its precision, recall, specificity, and accuracy.  As previously outlined in Section~\ref{sec:design}, the different operating modes of an LLM rater require achieving good performance in only a subset of the metrics we evaluate in order to be suitable for practical deployment.

\subsection{Ablation parameters}

\begin{figure}[t]
\centering

\begin{minipage}{\columnwidth}

   \begin{tcolorbox}[colback=green!5!white,
                    colframe=green!20!white,
                    arc=4mm,
                    ]
     \footnotesize {\bf \boldlab Misinformation and Disinformation Policy\boldrab}\\
1) Comments should not mislead census participants about the time, place, means, or eligibility requirements of the census.\\
2) Comments should not make false claims that could materially discourage census participation.\\
3) Comments should not mislead voters about the time, place, means, or eligibility requirements of voting.\\
4) Comments should not make false claims that could materially discourage voting.\\
5) Comments should not make false claims related to the technical eligibility requirements for current political candidates and sitting elected government officials to serve in office.\\
6) Comments should not encourage others to interfere with democratic processes.\\
7) Comments should not contain hacked info, the disclosure of which may interfere with democratic processes.\\
8) Comments should not make false claims that widespread fraud, errors, or glitches occurred in certain past elections to determine heads of government.\\
{\bf\boldlab/Misinformation and Disinformation Policy\boldrab}\\

{\bf Question:} You are an expert content moderator. Does the following comment violate the Misinformation and Disinformation Policy?\\

{\bf Comment: }``[COMMENT]''\\
{\bf Answer: }
  \end{tcolorbox}
\end{minipage}
\caption{Initial prompt derived from \Google's public policy around census and election misinformation. Across all our experiments, this policy language remains static.}
\label{fig:methods_prompt_example}
\end{figure}

\paragraph{Baseline policy specification} 

Based on early experimentation, we opted to seed all our prompt variants with a policy description derived from the harmful content policies shared publicly by \Google~\cite{google-policies}.  We reformatted all of these policies into an enumerated list that is easily interpretable by an LLM. We share an example in Figure~\ref{fig:methods_prompt_example} and provide the specification for all four policies we consider in the Appendix.

\paragraph{Zero-shot vs.\ few-shot} One variable we consider is whether an LLM requires examples of comments and verdicts to render a decision.  In our \emph{zero-shot} variant, we provide no examples but prepend a ``dummy'' policy and question-answer pair to prime the LLM on our expected output format (see Appendix). For our \emph{few-shot} variants, we provide a list of sample comments along with their human rated verdict, an example of which we show in Figure~\ref{fig:methods_prompt_with_example}. We opt for five examples: three comments that violate the policy under consideration and two that do not violate any policy.  

We explored three options for selecting these examples: (1) first, we used static, hand-picked examples.  We chose these examples to be representative of the overall problem space and the policy under question; e.g., for hate speech we chose violative examples that targeted different protected groups. After observing the success of this strategy (as discussed below in Section~\ref{sec:single}), we next experimented with (2) dynamically identifying semantically ``close'' comments in terms of their embedding distance; and (3) purposefully injecting errors into the dynamically chosen samples to understand the sensitivity of prompts to noise.

\paragraph{Keyword context and chain-of-thought} Another variable we consider is whether providing context on the most salient terms from violative comments assists an LLM with rendering an accurate verdict---a form of chain-of-thought~\cite{wei2022chain}. As part of our hand-picked few-shot examples, we include (manually chosen) keyword context as shown in Figure~\ref{fig:methods_prompt_with_example}.

\begin{figure}

\begin{minipage}{\columnwidth}

   \begin{tcolorbox}[colback=blue!5!white,
                    colframe=blue!20!white,
                    arc=4mm]
     \footnotesize 
     
{\bf Question:} You are an expert content moderator. Does the following comment violate the Misinformation and Disinformation Policy?\\

{\bf Comment:} "President Joe Biden hit them with Covid-19 to help him steal an election"\\
{\bf Answer:} Yes\\
{\bf Keywords:} covid-19 | steal | election\\
----\\
{\bf Comment: }``[COMMENT]''\\
{\bf Answer: }
  \end{tcolorbox}
\end{minipage}
\caption{A few-shot prompt variant that includes both an example policy-relevant comment and answer, and keyword context. The comment under evaluation appears after all examples.}
    \label{fig:methods_prompt_with_example}\vspace{-5pt}
\end{figure}

\paragraph{Model selection} 
Finally, we consider how different models affect performance.  We query all of our prompt variants using both the PaLM 2 \vertexsmall and \vertexlarge models (with the latter being larger than the former), available via Google's Vertex AI API~\cite{google-vertex-ai}, each with a \emph{context window} of 8192 tokens. For all experiments we use a fixed \emph{temperature} of zero, which means the LLM always picks the most probable next token, and generate one token for the LLM response. This determinism ensures that an LLM produces the same output across subsequent executions, which is critical for operating a classifier in practice and for evaluating errors that might otherwise be transient. We parse the LLM output, treating ``Yes'' (\ie 1) as a violative verdict and ``No'' (\ie 0) as non-violative verdict.  While we use only this binary output in this section, we discuss in Section~\ref{sec:deployment} the potential benefits of having the LLM output more information, such as a list of keywords and the policy entry that most contributed to its decision.

For privacy reasons, we were unable to evaluate using GPT-4 or Claude-2. Likewise, Gemini Pro and Ultra had not been released at the time of our experimentation. We emphasize that our goal is to assess the \emph{general} feasibility of content moderation via LLMs, rather than benchmarking the relative merit of different LLM providers. Additionally, individual LLM idiosyncrasies stemming from instruction tuning and reinforcement learning would likely require adjusting prompts per LLM architecture. Furthermore, it is cost prohibitive to query many models with all the prompt variants we evaluate in this study; as it is, with only two models we already perform half a million queries. 

\subsection{Single policy enforcement}
\label{sec:single}

\begin{table*}[t]
    \centering
    \input{tables/ablation_simplified}
    \caption{Performance of \vertexsmall and \vertexlarge at detecting harmful content in English. We report on three prompt variants: our zero-shot variant (acting as the baseline and containing just the policy language), our few-shot variant with hand-picked examples, and our hand-picked few-shot variant with keyword context. P-values indicate whether the differences between a variant and the baseline are statistically significant.}
    \label{table:ablation_single_policy}
\end{table*}

We begin by examining the accuracy of LLM raters that are specialized to enforce only a single policy at a time, evaluating each LLM rater on 10,000 
comments that violate the associated policy, and 10,000
comments that violate no policy. We present the results of our zero-shot prompt, few-shot prompt with hand-picked examples, and the hand-picked few-shot prompt with keyword context in Table~\ref{table:ablation_single_policy}, for both the models we used. (We evaluate the other two few-shot variants in subsequent experiments.) 

Compared to human raters, the best prompt and model combination achieves 98.7\% accuracy for election misinformation, 90.3\% accuracy for hate speech, 89.3\% accuracy for violent extremism, and 87.2\% accuracy for harassment. 

\paragraph{Comparing models} To compare the performance of models, we calculate a McNemar's test statistic comparing the best performing prompt for both \vertexlarge and \vertexsmall as evaluated on the same test set. Overall, \vertexlarge outperformed \vertexsmall, with a relative lift of 0.2\%--2.4\% between the best prompts ($p < 0.001$, except for hate speech where $p=0.177$). This modest performance gap indicates that platforms can opt for smaller and less costly models for some policies (especially keeping in mind that \vertexsmall is an order of magnitude less expensive than \vertexlarge, costing \$0.00025 per thousand input characters as compared to \$0.0025), but that the scaling laws of LLMs~\cite{kaplan2020scaling} still favor larger models for accurate policy interpretation.

\paragraph{Comparing prompt variants per model} To compare the performance of prompt variants, we calculate a McNemar's test statistic with the zero-shot prompt as the baseline. We find that our few-shot prompt yielded statistically equivalent or better performance overall for both \vertexsmall and \vertexlarge, with the exception of harassment where hand-picked examples sent to \vertexlarge degraded accuracy by 1\% ($p < 0.001$).  On the other hand, examples carry an additional cost in terms of increasing the length of the prompt (in the case of violent extremism, which has the shortest policy specification, by as much as 61\%) and thus the cost to evaluate the comment.

Keyword context appears unnecessary for \vertexlarge, but does yield a 6.2\% gain for violent extremism, and a 1.1\% gain for harassment for \vertexsmall $(p < 0.001)$. At the same time, keyword context lifted the recall of all policies with \vertexlarge---at the cost of precision. This suggests that keyword context can be omitted for LLM raters designed to automate decision where accuracy is more important than recall. Keywords may still play a role for smaller models.

\begin{figure}[t]
     \centering
     \includegraphics[width=\columnwidth]{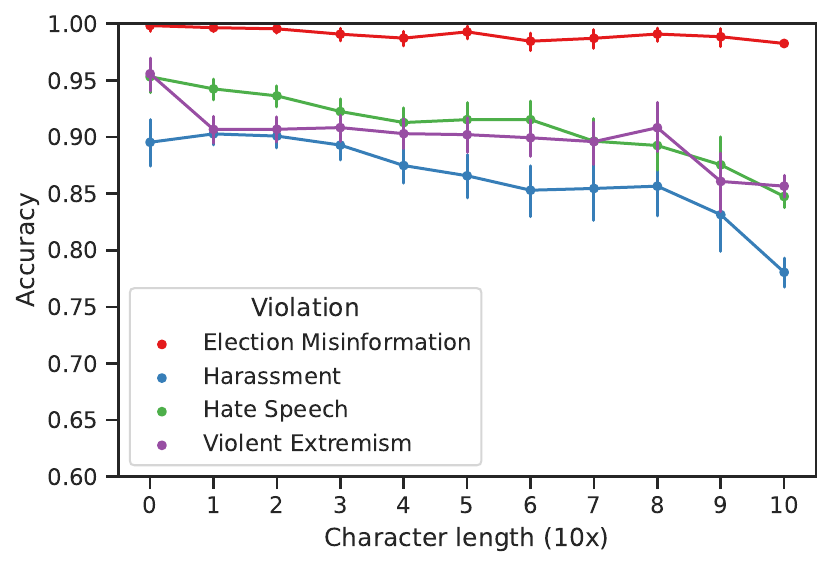}
    \caption{Accuracy of \vertexlarge for our hand-picked few-shot prompt variant. We segment our evaluation corpus into buckets of 0--9 characters, 10--19 characters, and so on up to 100+ characters. For each sample, we display error margins for a confidence level of 95\%. (Note the truncated Y-axis.)}
    \label{fig:text_length}
\end{figure}

\paragraph{Accuracy vs.\ content length} A final dimension we explore is the effect of longer, possibly extraneous text on classification accuracy. Of the comments in our dataset, 50\% have 11 or fewer words (and 59 characters) and 90\% have 51 or fewer words (and 282 characters). We present a breakdown of accuracy for an LLM rater configured to use \vertexlarge with our few-shot prompt across policies, stratified by character length in Figure~\ref{fig:text_length}. We find that our LLM rater declines in accuracy for longer comments. This may be a consequence of LLMs not giving enough attention to short violative passages in comments that are otherwise innocuous~\cite{liu2023lost}. A potential solution would be to segment long-form text and query individual passages, but this is likely prohibitively expensive if the prompt instructions remain longer than the text under evaluation.

\begin{table*}[t]
    \centering
    \input{tables/rag}
    \caption{Performance of our few-shot variant with examples chosen dynamically based on semantic distance from previously rated content. As a baseline, we use the hand-picked few-shot variant.  This prompt variant yielded our best performance across all variants, apart from election misinformation.}
    \label{table:rag}
\end{table*}

\subsection{Dynamic few-shot variants}

Given that our few-shot prompt variant with hand-picked examples reliably outperforms the zero-shot variant, we next examine how the quality of examples---in terms of both relevance and label accuracy---affects classification accuracy. Similarly to retrieval augmented generation (RAG)~\cite{lewis2020retrieval}, we dynamically generate a prompt for a given comment that includes five semantic nearest neighbors.  

For semantic similarity, we use {\small \sffamily textembedding-gecko} to produce 768-dimensional vector embeddings and an Annoy index~\cite{annoy} with 200 trees for fast, top-N approximate nearest-neighbor selection, with one index for 10,000 violative samples and one index for 10,000 non-violative samples per policy under evaluation. When querying these indices for the neighbors of a comment, we exclude the comment under evaluation as a valid neighbor. We then pick the three closest violations in our Annoy index, and the two closest non-violations. 

We present our results in Table~\ref{table:rag}, comparing to the performance of our hand-picked few-shot variant with a McNemar's test statistic. For \vertexlarge, we observe a performance change between -0.5\% and 3.9\% ($p < 0.001$).  Dynamic examples yielded the best overall performance across all prompt variants, other than for election misinformation. For \vertexsmall, dynamic examples hampered performance for election information (-5.0\%) and hate speech (-0.9\%), but improved accuracy for violent extremism (3.1\%) and harassment (2.1\%). 

That said, we find the performance of dynamic few-shot examples is sensitive to inaccurate labels. As an experiment, we selected the same semantically close comments but randomly flipped the verdict of one example to simulate a labeling error. We present the impact of these errors in Table~\ref{table:rag_bitlip}, again comparing to our hand-picked few-shot variant with a McNemar's test statistic. For \vertexlarge, hate speech detection degraded by 1.3\%, while election misinformation detection degraded by 4.6\% (all $p < 0.001$). We find \vertexsmall was more sensitive to noise, with performance degrading by as much as 7.2\% ($p < 0.001$). Harassment stands out for both model sizes, where noisy labels yielded better results than our hand-picked examples but worse accuracy than our zero-shot variant for \vertexlarge, or accurate dynamic labels.

Both of our experiments suggest that dynamically selecting few-shot examples can boost performance and potentially account for emerging concepts (\eg glorifying the name of a violent extremist, or a new hate term) not present in the training corpus of an LLM or the policy body of a prompt. However, queue operators need to ensure that examples favor high-quality labeled data to minimize any performance loss.

\begin{table}[t]
    \centering
    \input{tables/rag_bitflip_short}
    \caption{Performance for our few-shot variant with dynamic examples, but where we intentionally introduce a random verdict error for one example. As a baseline, we use the hand-picked few-shot variant. Both models are sensitive to errors, though losses in accuracy were lower for \vertexlarge. 
    }
    \label{table:rag_bitlip}
\end{table}

\subsection{Multi-policy enforcement}
While our focus thus far has been on how to optimize LLM raters for individual policies, in practice, platforms have tens of policy categories that need to be assessed against any object. As an alternative to sending content to $N$ specialized LLM raters, we explore the feasibility of deploying one zero-shot prompt encompassing multiple policies. To do this, we compile all our policy text (totalling 40 clauses) into a single prompt. We then compare this accuracy against our zero-shot, per-policy prompts on the same test corpus and calculate a McNemar's test statistic. We present our results in Table~\ref{tab:multipolicy}. We find our multi-policy prompt performs worse for both \vertexsmall (-2.6--10.9\%) and \vertexlarge (-2.2--13.7\%). The most impacted policy was election misinformation, likely due to it appearing at the end of the prompt's context window.

As such, prompts need to control for the \emph{verbosity} of policy language. Overly long policies will likely see a degradation in classification accuracy. That said, there are cost tradeoffs.  For a transformer-based model, the processing cost is $O(n^2 * m)$ where $n$ is the input length and $m$ is the output length (which is fixed at 1 for binary answers), which favors short, specialized prompts. However, other factors such as the base cost of running an inference and the size of the prompt may favor multi-policy prompts.  As a concrete estimate using the Google Cloud costs presented in Section~\ref{sec:performance}, the average character length of the comments in our dataset, and the four policies we consider, the cost of using \vertexlarge to evaluate a queue of 50,000 comments is roughly \$762 using a multi-policy prompt and roughly \$1,137 using single-policy few-shot prompts (in which each comment is fed to one LLM rater per policy).  We stress that these are estimates and that there are other ways to optimize prompts for cost.

\begin{table}[t]
    \centering
    \input{tables/multipolicy_acc_only}
    \caption{Performance for a single prompt covering all four content safety policies, with its 40 clauses ordered as hate speech, harassment, violent extremism, and election misinformation. Policies towards the end of the context window see a drop in accuracy compared to a prompt only related to that policy.}
    \label{tab:multipolicy}
\end{table}

\subsection{Summary}

Across all metrics, the overall best choice of model was \vertexlarge (the larger model).
In terms of our prompt variants, we generally saw the best performance from the few-shot variant with dynamically selected examples (with the caveats mentioned previously about cost and needing to ensure the accuracy of these examples).  As such, for the rest of the paper we consider only the results of querying \vertexlarge with this prompt variant.  We acknowledge, however, that models and prompt engineering practices are evolving rapidly, and given our focus on the general feasibility of using LLMs for content moderation we did not experiment with more advanced approaches like automated prompt optimization (APO)~\cite{zhang2023tempera,zhou2023large,pryzant2023automatic}.  We thus leave an exploration of the effectiveness of this, other practices, and even larger models as interesting future work.

%% file: tables/ablation_simplified.tex
\small
\begin{tabular}{l|c|cc|R{1.5cm}R{1.5cm}R{1.5cm}R{1.5cm}|rr}
\bf Policy & \bf Model & 
\rotatebox{45}{\bf Examples} & 
\rotatebox{45}{\bf Keywords} & 
{\bf Precision} &
{\bf Recall} & 
{\bf Specificity} & 
{\bf Accuracy} &
{\bf $\Delta$} &
{\bf p-value} \\
\toprule

\multirow{6}{*}{Hate} & \multirow{3}{*}{\vertexsmallshort} & 
\emptycirc &\emptycirc &86.0\% & \bf 95.1\% & 84.5\% & 89.8\% & -- & --\\
 & & \fullcirc & \emptycirc & \bf 86.6\% & 94.8\% & \bf 85.3\% & \bf 90.1\% & 0.3\% & 0.003\\
 & & \fullcirc & \fullcirc & 86.4\% & 94.7\% & 85.0\% & 89.9\% & 0.1\%& 0.336\\
 \cmidrule{2-10}
& \multirow{3}{*}{\vertexlargeshort} & 
\emptycirc &\emptycirc &85.5\% & \bf 96.6\% & 83.6\% & 90.1\% & -- & --\\
 & & \fullcirc & \emptycirc &86.4\% & 95.5\% & \bf 85.0\% & \bf 90.3\% &  0.2\%& 0.090\\
 & & \fullcirc & \fullcirc & \bf 86.5\% & 95.6\% & \bf 85.0\% & \bf 90.3\% & 0.2\%& 0.026\\
\midrule

\multirow{6}{*}{\shortstack[l]{Violent\\Extremism}} & \multirow{3}{*}{\vertexsmallshort} & 
 \emptycirc &\emptycirc &\bf 90.9\% & 68.2\% & \bf 93.2\% & 80.7\% & --& --\\
 & & \fullcirc & \emptycirc &88.6\% & 81.8\% & 89.5\% & 85.6\% & 5.0\%& $<$ 0.001\\
 & & \fullcirc & \fullcirc & 87.9\% & \bf 85.5\% & 88.2\% & \bf 86.9\% & 6.2\%& $<$ 0.001\\
\cmidrule{2-10} 
 & \multirow{3}{*}{\vertexlargeshort} & 
\emptycirc &\emptycirc &89.4\% & 86.6\% & \bf 89.7\% & 88.1\% & -- & --\\
 & & \fullcirc & \emptycirc & \bf 89.6\% &  88.9\% & \bf 89.7\% & \bf 89.3\%  & 1.2\%& $<$ 0.001\\
 & & \fullcirc & \fullcirc & 87.8\% & \bf 91.3\% & 87.3\% & \bf 89.3\% & 1.2\%& $<$ 0.001\\
 \midrule

\multirow{6}{*}{Harassment} & \multirow{3}{*}{\vertexsmallshort} & 
\emptycirc &\emptycirc & \bf 82.3\% & 85.9\% & \bf 81.5\% & 83.7\% & -- & --\\
 & & \fullcirc & \emptycirc &81.9\% & 87.2\% & 80.7\% & 84.0\% & 0.3\%& 0.123\\
 & & \fullcirc & \fullcirc & 81.9\% & \bf 89.2\% & 80.3\% & \bf 84.8\% & 1.1\%& $<$ 0.001\\
 \cmidrule{2-10} 
   & \multirow{3}{*}{\vertexlargeshort} & 
\emptycirc &\emptycirc & \bf 82.9\% & 93.7\% & \bf 80.7\% & \bf 87.2\% & -- &  --\\
 & & \fullcirc & \emptycirc &81.0\% & 94.6\% & 77.8\% & 86.2\% & -1.0\%& $<$ 0.001\\
 & & \fullcirc & \fullcirc & 80.1\% & \bf 95.9\% & 76.2\% & 86.1\% & -1.1\%& $<$ 0.001\\
   \midrule

 \multirow{6}{*}{\shortstack[l]{Election\\Misinformation}} & \multirow{3}{*}{\vertexsmallshort} & 
\emptycirc &\emptycirc    & \bf 96.1\% & 96.1\% & \bf 96.1\% & 96.1\% & -- & -- \\
 & & \fullcirc & \emptycirc & 95.7\% & \bf 99.4\% & 95.5\% & \bf 97.5\% & 1.3\%& $<$ 0.001\\
 & & \fullcirc & \fullcirc & 95.5\% & 99.3\% & 95.4\% & 97.3\% & 1.2\%& $<$ 0.001\\
  \cmidrule{2-10} 
    & \multirow{3}{*}{\vertexlargeshort} & 
\emptycirc &\emptycirc &98.2\% & 98.4\% & 98.2\% & 98.3\% & -- & --\\
 & & \fullcirc & \emptycirc &\bf 98.4\% & \bf 99.0\% & \bf 98.4\% &\bf  98.7\% & 0.4\%& $<$ 0.001\\
 & & \fullcirc & \fullcirc & 97.8\% & \bf 99.0\% & 97.8\% & 98.4\% & 0.1\%& 0.370\\
 \bottomrule
\end{tabular}

%% file: tables/rag.tex
\small
\begin{tabular}{l|c|rrrr|rr}
\toprule
\bf Policy & \bf Model & 
{\bf Precision} &
{\bf Recall} & 
{\bf Specificity} & 
{\bf Accuracy} &
\bf $\Delta$ &
\bf p-value \\

\midrule
  \multirow{2}{*}{Hate Speech} 
   & \vertexsmallshort & 83.3\% & 97.8\% & 80.4\% & 89.1\%& -0.9\% & $<$ 0.001\\
   & \vertexlargeshort & 88.2\% & 94.8\% & 87.3\% & 91.1\%& 0.8\% & $<$ 0.001\\
\midrule
  \multirow{2}{*}{Violent Extremism} & \vertexsmallshort & 84.4\% & 95.2\% & 82.3\% & 88.8\%& 3.1\%& $<$ 0.001\\
   & \vertexlargeshort & 92.2\% & 89.7\% & 92.5\% & 91.1\%& 1.8\%& $<$ 0.001\\
\midrule
  \multirow{2}{*}{Harassment} & \vertexsmallshort & 79.0\% & 98.0\% & 74.0\% & 86.0\%& 2.1\%& $<$ 0.001\\
  & \vertexlargeshort & 85.7\% & 96.4\% & 83.9\% & 90.1\%& 3.9\%& $<$ 0.001\\
\midrule
 \multirow{2}{*}{Election Misinformation} & \vertexsmallshort & 87.1\% & 99.7\% & 85.3\% & 92.5\%& -5.0\%& $<$ 0.001\\
 & \vertexlargeshort & 99.0\% & 97.4\% & 99.1\% & 98.2\%& -0.5\%& $<$ 0.001\\
 \bottomrule
\end{tabular}

%% file: tables/rag_bitflip_short.tex
\small
\begin{tabular}{p{2cm}|c|r|rr}
\toprule
\bf Policy & \bf Model & 
{\bf Accuracy} &
\bf $\Delta$ &
\bf p-value \\

\midrule

Hate &\vertexsmallshort & 88.4\% & -1.7\%& $<$ 0.001\\
Speech  & \vertexlargeshort & 88.9\%& -1.3\%& $<$ 0.001\\
   \midrule
Violent & \vertexsmallshort & 86.6\%& 0.9\%& $<$ 0.001\\
Extremism & \vertexlargeshort & 86.6\%& -2.7\%& 0.023\\
 \midrule
\multirow{2}{*}{Harassment} & \vertexsmallshort & 84.7\%& 0.7\%& $<$ 0.001\\
& \vertexlargeshort & 86.8\%& 0.6\%& $<$ 0.001\\
\midrule
Election & \vertexsmallshort & 90.3\%& -7.2\%& $<$ 0.001\\
Misinformation & \vertexlargeshort & 94.0\%& -4.6\%& $<$ 0.001\\
 \bottomrule
\end{tabular}

%% file: tables/multipolicy_acc_only.tex
\small
\begin{tabular}{l|c|r|rr}
\toprule
\bf Policy & \bf Model & 
{\bf Accuracy} &
\bf $\Delta$ &
\bf p-value\\
\midrule
Hate & \vertexsmallshort & 86.3\%& -3.4\%& $<$ 0.001\\
Speech & \vertexlargeshort & 87.5\%& -2.6\%& 0.002\\
\midrule
Violent&  \vertexsmallshort & 77.3\%& -3.4\%& $<$ 0.001\\
extremism   & \vertexlargeshort & 79.0\%& -9.2\%& 0.023\\
\midrule
  \multirow{2}{*}{Harassment} & \vertexsmallshort & 81.1\%& -2.6\%& $<$ 0.001\\
  & \vertexlargeshort & 85.0\%& -2.2\%& 0.055\\
\midrule
Election & \vertexsmallshort & 85.2\%& -10.9\%& $<$ 0.001\\
Misinformation & \vertexlargeshort & 84.6\%& -13.7\%& $<$ 0.001\\
 \bottomrule
\end{tabular}

%% file: 04a_patterns.tex
\section{Utilizing LLM raters}
Based on the findings of our ablation study, Section~\ref{sec:experiments}, we explore the suitability of our most accurate prompt variant---\vertexlarge with dynamic few-shot examples---in satisfying each of the collaborative design patterns outlined earlier in Section~\ref{sec:design}. As part of this, we introduce a technique that allows a single prompt to flexibly adapt to every design pattern.

\begin{figure}[t]
    \centering
    \includegraphics[width=\columnwidth]{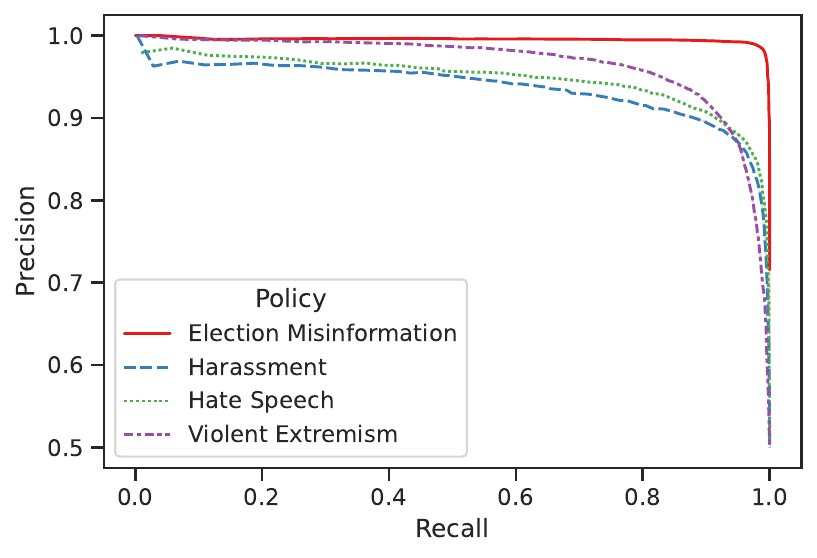}
    \caption{Precision vs. recall for our adaptive few-shot example policy prompt and \vertexlarge. Using the probability of tokens predicted by an LLM, we can flexibly alter whether a prompt favors precision or recall.}
    \label{fig:score_threshold_pr_curve}
\end{figure}

\subsection{Dynamically tuning LLM raters} Our experiments show that one prompt variant might yield the best precision, while another yields the best recall or specificity. In practice, maintaining multiple variants per policy---especially at the speed that policies might adapt---is prohibitively expensive. Instead, we examine how to use the  probabilities generated by LLMs (\eg top-k scores for returned tokens) to create a configurable decision boundary to support multiple use cases without needing to change the underlying prompt. Using \vertexlarge with a dynamic few-shot example prompt, we configure the model to return {\sf \small p=Score(``Yes'')}; i.e., the probability
that the next generated token after our prompt is ``Yes''.\footnote{In experimentation, {\sf \footnotesize p=Score(``No'')} was equally viable, though we note the two are not guaranteed to add to 1.0.} From this probability, we define a sliding decision threshold $T$, where we treat all $p \ge T$ as a policy violation and all other samples as non-violative. We then calculate the precision and recall at each threshold $T$.

We present the precision and recall curve of this approach in Figure~\ref{fig:score_threshold_pr_curve}. Misinformation stands out as its overall accuracy is 98.2\%, while the other policies have an accuracy of 90.1--91.1\%. The token probabilities returned by an LLM can effectively be interpreted as a confidence score. Comments that receive a ``Yes'' verdict can still be treated as non-violative if the LLM does not meet the requisite confidence threshold $T$. For scenarios where recall is imperative, $T < 0.5$ is most suitable; for precision, $T > 0.5$ is most suitable.

\subsection{Collaborative design pattern performance}
We simulate the performance of each collaborative design pattern---previously discussed in Section~\ref{sec:design}---assuming a single rating queue per policy that contains a 1:1 mixture of violative and non-violative content.

\paragraph{Pre-filtering} A pre-filtering LLM rater aims to remove clearly non-violative content from a rating queue, while falling back on human rater expertise for borderline or egregious content. An LLM rater could remove 54.5--98.2\% of comments from our simulated rater queue due to being non-violative, while ensuring 99\% of violative content is sent to a human rater (Table~\ref{table:score_threshold_filtering}). If constraints are relaxed and the LLM rater need only achieve a recall of 95\%, then an LLM rater could remove 85.2--99.3\% of non-violative comments from our simulated queues.

\begin{table}[t]
    \centering
    \input{tables/score_threshold_specificity}
    \caption{Performance of an LLM rater pre-filtering non-violative comments with minimum recall threshold of 95\% and 99\%. The higher variance in the pre-filtering rate for 99\% recall may be due to label inaccuracy, where the LLM is highly confident a ground truth violation is non-violative, requiring a very low threshold $T$ that passes too much content on to a human rater.}
    \label{table:score_threshold_filtering}
\end{table}

\paragraph{Rapid escalation} A rapid escalation LLM rater aims to escalate clearly violative content without requiring verification by a human rater, while falling back to human raters on all other content. An LLM rater could automatically escalate 48.8--99.7\% of violative content with a precision of 95\% (Table~\ref{table:score_threshold_escalation}). A precision threshold greater than 99\% is not feasible for hate speech and harassment, likely due to our ground truth incorrectly labeling some hate speech and harassment as non-violative. For the remaining policies, an LLM rater could automatically escalate 41.5--97.4\% of violative content.

\begin{table}[t]
    \centering
    \input{tables/score_threshold_precision}
    \caption{Performance of an LLM rater rapidly escalating violative content with minimum precision of 95\% and 99\%. An N/A means the targeted precision is currently not achievable on our dataset.}
    \label{table:score_threshold_escalation}
\end{table}

\paragraph{Autonomous rating} An autonomous LLM rater aims to handle ratings without any fallback to human raters. In practice, such an LLM rater achieves only 90.1--98.2\% of the accuracy of human raters (Table~\ref{table:score_threshold_accuracy}). As such, the aforementioned collaborative design patterns may be more suitable given the current capabilities of LLMs. 

\begin{table}[t]
    \centering
    \input{tables/score_threshold_accuracy}
    \caption{Performance of autonomous rating for our adaptive few-shot example policy prompt and \vertexlarge.}
    \label{table:score_threshold_accuracy}
\end{table}

\paragraph{Human rater validation} Our validation LLM rater aims to improve overall decision making accuracy by identifying likely human errors. For our simulated queues, we reflect on the 80,000 rating decisions an autonomous LLM rater would have made and compare them to our ground truth labels from human raters. Our LLM rater disagreed with human raters on 5,896 comments, or 7.4\% of the time. We randomly sampled 400 points of disagreement, stratifying this selection to include 100 comments for each simulated policy rating queue. We then manually reviewed each comment using the same methodology as in Section~\ref{sec:methods_validation} to determine whether researchers agreed with the policy expert or the LLM rater. We treat these as \emph{tie-breaking ratings}.\footnote{As with our earlier validation, we caution researchers are not trained policy experts, so concurrence with an LLM may be inaccurate.}

We present our results in Table~\ref{table:disagreement}. For hate speech, our tie-breaking rating agreed with LLMs 49\% of the time, versus the ground truth 43\% of the time. As an illustrative example, our LLM rater flagged ``nice [thumbs up emoji] 88'', a Nazi reference, which was potentially missed by the human rater. However, our LLM rater missed ``Cull these knickers'', a likely slur. For election misinformation, tie-breaking ratings matched those of human raters more often than LLMs---52\% versus 43\% respectively. The loss in accuracy here is due to our LLM rater overindexing on any form of misinformation (particularly COVID-19, but also conspiracy theories) which fell outside the stated policy. In other cases, our LLM rater missed a shorthand of misinformation like ``80 million dead people,'' a census-related claim that more people voted in 2020 than was possible. For violent extremism, tie-breaking ratings most often concurred with human raters (or lacked context), in part due to the policy more cautiously flagging any mention of terrorist groups, whereas the LLM appeared to apply more of a contextual lens.

Taken as a whole, our qualitative analysis suggests that pairing LLM raters with human raters can measurably improve the accuracy of decision making. It also demonstrates that our LLM raters may be more accurate than our ground truth indicates. In practice, queue operators can configure the number of tie-breaking ratings using a higher precision score threshold.

\begin{table}[]
    \centering
    \input{tables/disagreement}
    \caption{Breakdown of tie-breaking ratings for instances where an LLM rater disagreed with our ground truth. We indicate whether the tie-breaking vote agreed with the ground truth (human correct), with the LLM rater (LLM correct), or whether critical context was missing to render a tie-breaking rating.}
    \label{table:disagreement}
\end{table}

\paragraph{Human rater assistance} An LLM rater assistant aims to provide critical context to human raters as they render verdicts. As our dataset and analysis is retroactive, we defer evaluation of this design pattern to our real-world experiments.

%% file: tables/score_threshold_specificity.tex
\small
\begin{tabularx}{\columnwidth}{X|R{2cm}|R{2cm}}
 \toprule
\bf Policy & \bf Recall threshold & \bf Pre-filtering rate \\
\midrule
Hate Speech & 95.5\% & 86.6\% \\
Violent Extremism & 95.0\% & 85.8\% \\
Harassment & 95.5\% & 85.2\% \\
Election Misinformation & 95.1\% & 99.3\% \\
 \midrule
Hate Speech & 99.1\% & 75.1\% \\
Violent Extremism & 99.1\% & 54.5\% \\
Harassment & 99.1\% & 71.1\% \\
Election Misinformation & 99.0\% & 98.2\% \\
\bottomrule
\end{tabularx}

%% file: tables/score_threshold_precision.tex
 \small
 \begin{tabularx}{\columnwidth}{X|R{2cm}|R{2cm}}
 \toprule
\bf Policy & \bf Precision threshold & \bf Violations escalated \\
\midrule
Hate Speech & 95.2\% & 60.6\% \\
Violent Extremism & 95.1\% & 82.5\% \\
Harassment & 95.2\% & 48.8\% \\
Election Misinformation & 95.0\% & 99.7\% \\
\midrule
Hate Speech & 99.0\% & N/A \\
Violent Extremism & 99.0\% & 41.5\% \\
Harassment & 99.0\% & N/A \\
Election Misinformation & 99.0\% & 97.4\% \\
\bottomrule
\end{tabularx}

%% file: tables/score_threshold_accuracy.tex
 \small
 \begin{tabularx}{\columnwidth}{X|R{2cm}}
 \toprule
\bf Policy & \bf Accuracy \\
\midrule
Hate Speech  &  91.1\% \\
Violent Extremism & 91.1\% \\
Harassment & 90.1\% \\
Election Misinformation & 98.2\% \\
\bottomrule
\end{tabularx}

%% file: tables/disagreement.tex
\small
\begin{tabularx}{\columnwidth}{X|P{1cm}|P{1cm}|P{1cm}}
\toprule
\bf Policy & 
\rotatebox{0}{\makecell{\bf  Human \\ \bf correct}} & 
\rotatebox{0}{\makecell{\bf LLM \\ \bf correct}} & 
\rotatebox{0}{\makecell{\bf Missing \\ \bf context}} \\
\midrule
Hate Speech & 43\% & 49\% & 8\% \\
Violent Extremism & 50\% & 34\% & 16\% \\
Harassment & 47\% & 43\% & 10\% \\
Election Misinformation & 52\% & 43\% & 5\% \\
\bottomrule
\end{tabularx}

%% file: 05_deployment.tex
\section{LLM raters in the real world}\label{sec:deployment}
To demonstrate the feasibility of LLM raters, we piloted the integration an LLM rater into a live rating queue at Google containing English content. The LLM rater served two purposes: to optimize what content went to human raters and to assist human raters in arriving at a verdict.

\paragraph{Prompt design} The rating queue in our pilot triages user-generated reports of potentially harmful content related to 20+ policies. Roughly 90\% of these reports are inaccurate, with human raters ultimately determining content is non-violative. Due to engineering and cost complexity, we limited our design to a single, multi-policy prompt covering all of these policies. We used \vertexlarge for our model. Per our earlier analysis, we found \vertexlarge's accuracy degraded with long policies. We mitigated this by simplifying all of the policy clauses with an emphasis on recall. For example, a policy against dehumanizing a person based on their race, or stereotyping someone based on their race, might be up-leveled to a single statement against talking about a person's race whatsoever. Any false positives in a real deployment would be safeguarded by human raters. Conversely, quickly identifying errant user reports would better optimize human rater expertise.

\paragraph{Optimizing human rater expertise} We ran a pilot of our LLM rater in a shadow mode on a random sample of N=2,700 objects over a 30-day period, exclusively for English content. This non-enforcement ensured our LLM rater did not impact the quality of the rating queue, while allowing us to evaluate the efficacy of our proposed strategy. We sent every sampled object to both our LLM rater and the human rater process.\footnote{In practice, a single piece of content might be rated by multiple human raters and de-conflicted according to \Google's processes. We use the final verdict after any such processes.} Over this 30-day period, we estimate that for a fixed recall of 95.1\%, our LLM rater could reduce the volume of content requiring rating by 41.5\% (\eg pre-filtering design pattern).

\paragraph{Improving accuracy of verdicts} As a separate pilot, we configured our LLM rater to output not just a verdict, but also keywords explaining the verdict (\eg human rater assistance design pattern). As a within-subjects experiment to human raters who opted-in to the keyword feature, we surfaced these keywords for two weeks, by highlighting their appearance within the original text content rated by humans (N=500+ objects), comparing it against ratings made by the same human raters in the two week prior period conducted without support from the LLM (N=500+). Again, this was exclusively for English content. Recall for human raters---judged via existing quality control systems---increased by 9\% absolute ($p < 0.001$), while precision increased by 11\% absolute ($p < 0.001$). There was no statistically significant change in the time it took to review content.

%% file: 06_discussion.tex
\section{Discussion}\label{sec:discussion}
In light of our findings, we discuss the benefits of LLM raters, potential risks, optimization strategies for improving performance, and research opportunities for the future. 

\subsection{Benefits}

\paragraph{Scale \& maintenance} LLM raters can optimize available human expertise to a scale that was previously cost- and toil-prohibitive. While the current price per inference makes LLM raters best suited for rating content that is sent to existing human rater queues, in the future, LLM costs are likely to decrease. This points to a possibility of LLM raters acting as scaled classifiers. Such an architecture would minimize the training data required compared to existing classifiers, with LLM raters requiring only tens of samples for fixed few-shot prompts, hundreds of samples for adaptive few-shot prompts, or thousands of samples for fine-tuning.

\paragraph{Rapid response \& iteration} LLM raters enable policy experts to minimize the time between when a threat emerges and when a platform deploys a protection. Policy experts can distill their requirements into a simple set of instructions that an LLM rater can immediately enforce. This agility in adapting policy language---where policy experts rather than engineers can directly iterate on prompts---is critical to responding to the global landscape of user concerns, threat actors, and regulations. Paired with other scaled defense-in-depth systems, LLM raters allow for upstream detectors to be less accurate (and thus easier to develop and deploy), with LLM and human raters correcting for any errors before any enforcement action.

\paragraph{Consistency} Evolving policies, different cultural norms, and different levels of expertise mean that two human raters can arrive at different verdicts. This is particularly true when human raters are in the midst of adapting their processes to policy changes. LLM raters allow for more consistent---and potentially unbiased---enforcement, acting as a safety net.

\paragraph{Reducing emotional burden} LLM raters offer an opportunity to automatically identify the most egregious harmful content, removing the need for human rating. While human raters would still be involved in borderline cases, this shift in balance may improve the well-being of human raters~\cite{steiger2021psychological}.

\subsection{Risks}

\paragraph{Prompt injection \& padding} Similar to traditional security classifiers, LLM raters are potentially susceptible to evasion, particularly via prompt injection~\cite{perez2022ignore, abdelnabi2023not}. Addressing this requires future research into separating trusted and untrusted inputs and their ability to influence LLM decision making. Alternatively, LLM raters may miss short passages of violative content intentionally padded with otherwise benign content. Addressing this requires either segmenting passages or improving the attention mechanism of LLM raters.

\paragraph{Model drift \& idiosyncrasies} As new LLMs are regularly launched, prompts previously optimized for one version of a model may degrade in performance with the release of a new version. Likewise, optimizations for one LLM may not generalize to another LLM due to idiosyncrasies in RLHF tuning. Safeguarding against regressions requires maintaining both a prompt and evaluation dataset to benchmark changes.

\paragraph{Concept lag \& bias} The foundational models underpinning LLMs require massive compute resources which can lead to concept lag where an LLM is unaware of recent events. This poses a challenge for emerging threats---like a violent event or scam tactic---that may not appear in a foundational model's training data. This risk can potentially be mitigated via our dynamic few-shot example approach, supplying models with emerging context via examples.  Related, LLMs may be affected by stereotype or biases~\cite{nadeem2020stereoset} which can interfere with the detection of hate speech or harassment. Addressing this risk requires further improvements to debiasing LLMs~\cite{guo2022auto, lauscher2021sustainable, meade2021empirical}.

\subsection{Optimizations \& future work}

\paragraph{Optimizing LLM costs vs. performance} An inherent complexity with LLM raters is that the cost per inference depends on the size of the prompt and the size of the LLM. Our experiments demonstrate that platforms have a variety of ways to prioritize cost vs.\ accuracy. Strategies include using smaller, less expensive LLMs (\eg \vertexsmall vs. \vertexlarge); simplifying policies via automatic prompt optimization~\cite{pryzant2023automatic, yang2023large}; or optimizing the number of examples in a prompt or even the candidate selection strategy~\cite{li2022survey}. We envision that real-world deployments will consist of distinct layers of LLM raters each with supporting RAG databases. For instance, a platform may first query a fast-path, zero-shot, multi-policy LLM rater with high specificity and recall. Potential violations would then fan out to multiple, specialized, single-policy prompts with adaptive few-shot examples running on a larger LLM model, where any positive verdict from this final layer would result in an escalation.

\paragraph{Beyond platforms} While our work focused on removing harmful content from platforms, in practice, the ease of configuring prompts and score thresholds means that LLM raters can be equally useful tools for community moderation or even personalized content filters. The need for such tools stems from differing opinions on what constitutes harmful content, particularly in the context of hate and harassment~\cite{jhaver2023personalizing, gordon2022jury, kumar2021designing}.

\paragraph{Safety benchmark} Making forward progress as a research community towards developing LLM raters requires a canonical ``safety benchmark''. Such a dataset would encompass a multi-lingual, multi-modal testing corpus that spans a variety of content safety concepts (\eg scams, hate, misinformation, and more). Such a benchmark would provide a universal metric for judging progress of LLMs and prompts towards improving the state of user safety online. 

\paragraph{Other languages \& regional context}
Given the fluency of our researcher team and our focus on feasibility, our initial study used only comments in English. Further testing is needed for a wider selection of languages to understanding the performance of LLM raters for a global context.

%% file: 07_conclusions.tex
\section{Conclusions}
We investigated the feasibility of leveraging large language models (LLMs) to automate or otherwise assist human raters with identifying harmful content. Using a dataset of 50,000  
comments including hate speech, harassment, violent extremism, and election misinformation, we demonstrated that Google's PaLM 2 \vertexlarge can achieve over 90\% accuracy when compared to  human verdicts. We outlined a variety of design patterns that leveraged this detection capability, such as pre-filtering non-violative content, rapidly escalating violative content, or supporting humans with their decision making. We piloted our proposed approach and found we could automate 41.5\% of decisions for a live review queue, while  improving the precision and recall of human raters by 9--11\% (absolute). In summary, this paper shows an extremely promising start for integrating LLMs into the infrastructure that platforms use to protect users from harmful content.

%% file: 08_appendix.tex
\section*{Appendix}

\begin{figure*}[t]
\centering

\begin{minipage}{\textwidth}
   \begin{tcolorbox}[colback=pink!5!white,
                    colframe=pink!20!white,
                    arc=4mm]
   \footnotesize 
{\bf \boldlab Dangerous or Illegal Policy\boldrab}\\
1) Comments should not discuss minors drinking alcohol, using vaporizers, e-cigarettes, tobacco or marijuana.\\
2) Comments should not discuss bomb making.\\
3) Comments should not discuss how to to kill or harm others.\\
4) Comments should not discuss how to steal goods or promote dishonest behavior.\\
{\bf \boldlab/Dangerous or Illegal Policy\boldrab}\\

{\bf Question:} You are an expert content moderator. Does the following comment violate the Dangerous or Illegal Policy?\\
{\bf Comment: }"How to make a pipe bomb. You will need duct tape, alcohol, matches, a fuse, and black powder."\\
{\bf Answer: }Yes
\end{tcolorbox}
\end{minipage}
\caption{The dummy policy and example that we prepend to the real policy specification and comment in our zero-shot variant.  We do this to prime the expected output format; providing instructions to the LLM to only answer ``Yes'' or ``No'' is a viable alternative.}
\label{fig:dummy_policy}
\end{figure*} 

\begin{figure*}[t]
\centering

\begin{minipage}{	\textwidth}
  \begin{tcolorbox}[colback=green!5!white,
                    colframe=green!20!white,
                    arc=4mm]
  \footnotesize 
{\bf \boldlab Hate Speech Policy\boldrab}\\
1) Comments should not promote violence or hatred against individuals or groups based on age, caste, disability, ethnicity, gender identity and expression, nationality, race, immigration status, religion, sex/gender, sexual orientation, victims of a major violent event and their kin, or veteran status.\\
2) Comments should not encourage violence against individuals or groups based on any of the attributes noted above.\\
3) Comments should not incite hatred against individuals or groups based on any of the attributes noted above.\\
4) Comments should not dehumanize individuals or groups by calling them subhuman, comparing them to animals, insects, pests, disease, or any other non-human entity.\\
5) Comments should not praise or glorify violence against individuals or groups based on the attributes noted above.\\
6) Comments should not use racial, religious or other slurs and stereotypes that incite or promote hatred based on any of the attributes noted above. this can take the form of speech, text, or imagery promoting these stereotypes or treating them as factual.\\
7) Comments should not claim that individuals or groups are physically or mentally inferior, deficient, or diseased based on any of the attributes noted above. This includes statements that one group is less than another, calling them less intelligent, less capable, or damaged.\\
8) Comments should not allege the superiority of a group over those with any of the attributes noted above to justify violence, discrimination, segregation, or exclusion.\\
9) Comments should not contain conspiracy theories saying individuals or groups are evil, corrupt, or malicious based on any of the attributes noted above.\\
10) Comments should not call for the subjugation or domination over individuals or groups based on any of the attributes noted above.\\
11) Comments should not deny or minimize a well-documented, major violent event or the victimhood of such an event.\\
12) Comments should not attack a person's emotional, romantic and/or sexual attraction to another person.\\
13) Comments should not contain hateful supremacist propaganda including the recruitment of new members or requests for financial support for their ideology.\\
{\bf \boldlab/Hate Speech Policy\boldrab}\\

{\bf Question:} You are an expert content moderator. Does the following comment violate the Hate Speech Policy?\\
{\bf Comment: }"[COMMENT]"\\
{\bf Answer:}
\end{tcolorbox}
\end{minipage}
\caption{Zero-shot hate speech detection prompt with policy language derived from \Google's published policy on Hate Speech~\cite{google-policies}.}
\label{fig:prompt_example}
\end{figure*} 

\begin{figure*}[t]
\centering

\begin{minipage}{	\textwidth}
  \begin{tcolorbox}[colback=green!5!white,
                    colframe=green!20!white,
                    arc=4mm]
  \footnotesize 
{\bf \boldlab Violent Extremism Policy\boldrab}\\
1) Comments should not incite others to commit violent acts against individuals or a defined group of people.\\
2) Comments should not promote, recruit, radicalize, or coordinate violence, murder, mass shootings, organized crime, or terrorism.\\
3) Comments should not glorify or celebrate violence, extremism, or terrorist organizations like Hezbolla, ISIS, or the IRA.\\
4) Comments should not glorify or celebrate mass shootings or the shooters or attackers involved.\\
5) Comments should not promote violent uprising against a government\\
6) Comments should not discuss violence involving minors.\\
7) Comments should not discuss violent physical sexual assault.\\
8) Comments should not discuss malicious mistreatment of animals or how to cause animals to experience distress.\\
9) Comments should not glorify or promote serious neglect, mistreatment, or harm toward animals.\\
{\bf \boldlab/Violent Extremism Policy\boldrab}\\

{\bf Question:} You are an expert content moderator. Does the following comment violate the Violent Extremism Policy?\\
{\bf Comment: }"[COMMENT]"\\
{\bf Answer:}
\end{tcolorbox}
\end{minipage}
\caption{Zero-shot violent extremism detection prompt with policy language derived from \Google's published policy on Violent Extremism~\cite{google-policies}.}
\label{fig:prompt_example}
\end{figure*} 

\begin{figure*}[t]
\centering

\begin{minipage}{	\textwidth}
  \begin{tcolorbox}[colback=green!5!white,
                    colframe=green!20!white,
                    arc=4mm]
  \footnotesize 
{\bf \boldlab Harassment Policy\boldrab}\\
1) Comments should not include name calling or malicious insults (such as racial slurs) based on someone's intrinsic attributes. These attributes include their protected group status, physical attributes, or their status as a survivor of sexual assault, non-consensual intimate imagery distribution, domestic abuse, child abuse and more.\\
2) Comments should not shame, deceive or insult a minor. A minor is defined as an individual under the legal age of majority. This usually means anyone younger than 18 years old, but the age of a minor might vary by geography.\\
3) Comments should not reveal someone’s personally identifiable information (PII), such as their home address, email addresses, sign-in credentials, phone numbers, passport number, medical records, or bank account information.\\
4) Comments should not incite others to harass or threaten individuals.\\
5) Comments should not encourage doxxing, dogpiling, brigading or off-platform targeting.\\
6) Comments should not claim someone is part of a harmful conspiracy theory where the conspiracy theory has been linked to direct threats or violent acts.\\
7) Comments should not make implicit or explicit threats of physical harm or destruction of property against identifiable individuals. Implicit threats include threats that don’t express a specific time, place or means, but may feature weapon brandishing, simulated violence and more.\\
8) Comments should not revel in or mocking the death or serious injury of an identifiable individual.\\
9) Comments should not discuss violence against others (executions, torture, maimings, beatings and more).\\
10) Comments should not include unwanted sexualization or anything that graphically sexualizes or degrades an individual.\\
{\bf \boldlab/Harassment Policy\boldrab}\\

{\bf Question:} You are an expert content moderator. Does the following comment violate the Harassment Policy?\\
{\bf Comment: }"[COMMENT]"\\
{\bf Answer:}
\end{tcolorbox}
\end{minipage}
\caption{Zero-shot harassment detection prompt with policy language derived from \Google's published policy on Harassment~\cite{google-policies}.}
\label{fig:prompt_example}
\end{figure*} 

\begin{figure*}[t]
\centering

\begin{minipage}{	\textwidth}
  \begin{tcolorbox}[colback=green!5!white,
                    colframe=green!20!white,
                    arc=4mm]
  \footnotesize 
{\bf \boldlab Misinformation and Disinformation Policy\boldrab}\\
1) Comments should not mislead census participants about the time, place, means, or eligibility requirements of the census.\\
2) Comments should not make false claims that could materially discourage census participation.\\
3) Comments should not mislead voters about the time, place, means, or eligibility requirements of voting.\\
4) Comments should not make false claims that could materially discourage voting.\\
5) Comments should not make false claims related to the technical eligibility requirements for current political candidates and sitting elected government officials to serve in office.\\
6) Comments should not encourage others to interfere with democratic processes.\\
7) Comments should not contain hacked info, the disclosure of which may interfere with democratic processes.\\
8) Comments should not make false claims that widespread fraud, errors, or glitches occurred in certain past elections to determine heads of government.\\
{\bf \boldlab/Misinformation and Disinformation Policy\boldrab}\\

{\bf Question:} You are an expert content moderator. Does the following comment violate the Misinformation and Disinformation Policy?\\
{\bf Comment: }"[COMMENT]"\\
{\bf Answer:}
\end{tcolorbox}
\end{minipage}
\caption{Zero-shot election misinformation detection prompt with policy language derived from \Google's published policy on Census and Election Misinformation~\cite{google-policies}.}
\label{fig:prompt_example}
\end{figure*} 